\documentstyle[12pt,epsf]{article}
\begin{document}
\newcommand{\bra}{\langle}
\newcommand{\ket}{\rangle}
\newcommand{\eq}[2]{\begin{equation}\label{#1} #2 \end{equation}}
\newcommand{\tbf}[1]{{\bf #1}}
\newcommand{\tit}[1]{{\it #1}}
\newcommand{\cl}[1]{{\cal #1}}
\newcommand{\half}{{\scriptstyle \frac12}}
\newcommand{\intRR}{\int\limits_{-\infty}^{\infty}}
\newcommand{\intR}{\int\limits_{0}^{\infty}}
\newcommand{\LRD}[1]{\frac{{{\displaystyle\leftrightarrow}
\atop {\displaystyle\partial}}}{\partial #1}}
\newcommand{\diff}[1]{\partial /{\partial #1}}
\newcommand{\Diff}[1]{\frac{\partial}{\partial #1}}
\newcommand{\DiffM}[2]{\frac{\partial #1}{\partial #2}}
\newcommand{\HDiff}[2]{\frac{\partial^{#1}}{\partial #2^{#1}}}
\newcommand{\hc}[1]{{#1}^{\dagger}}
\newcommand{\disp}{\displaystyle}
\newcommand{\inn}{{\rm in}}
\newcommand{\out}{{\rm out}}
\newcommand{\vac}{{\rm vac}}
\newcommand{\intT}{\int\limits_{-\infty}^{t}}
\newcommand{\intXI}{\int\limits_0^{\xi}}
\newcommand{\aop}{{\rm a}}
\newcommand{\hyp}{{}_2 F_1}
\newcommand{\multeq}[2]{\eq{#1}{\begin{array}{l} \displaystyle
#2 \end{array}}}
\newcommand{\el}{ \\ \\ \displaystyle}
\newcommand{\then}{\Rightarrow}
\newcommand{\da}{\downarrow}
\newcommand{\ua}{\uparrow}
\renewcommand{\Re}{\,{\rm Re}\,}

\title{
\vspace{-2cm} "Shaking" of an atom in a non-stationary cavity }
\author{
A.M. Fedotov\thanks{fedotov@cea.ru}, N.B. Narozhny\thanks{narozhny@pc1k32.mephi.ru}\\
{\small{\it Moscow State Engineering Physics Institute}} \\ {\small{\it (Technical
University), Moscow 115409, Russia}} \\ and Yu.E.
Lozovik\thanks{lozovik@isan.troitsk.ru}
\\ {\small{\it Institute of Spectroscopy of Russian Academy of Science}},\\
{\small{\it Troitsk 142090, Russia}}}
\date{\vphantom{1.5cm}}
\maketitle
\vspace{-2cm}

\abstract {\small{ We consider an atom interacting with a quantized electromagnetic
field inside a cavity with variable parameters. The atom in the ground state located
in the initially empty cavity can be excited by variation of cavity parameters. We
have discovered two mechanisms of atomic excitation. The first arises due to the
interaction of the atom with the non-stationary electromagnetic field created by
modulation of cavity parameters. If the characteristic time of variation of cavity
parameters is of the order of the atomic transition time, the processes of photon
creation and atomic excitation are going on simultaneously and hence excitation of
the atom cannot be reduced to trivial absorption of the photons produced by the
dynamical Casimir effect. The second mechanism is "shaking" of the atom due to fast
modulation of its ground state Lamb shift which takes place as a result of fast
variation of cavity parameters. The last mechanism has no connection with the vacuum
dynamical Casimir effect. Moreover, it opens a new channel of photon creation in the
non-stationary cavity. Nevertheless, the process of photon creation is altered by the
presence of the atom in the cavity, even if one disregards the existence of the new
channel. In particular, it removes the restriction for creation of only even number
of photons and also changes the expectation value for the number of created photons.
Our consideration is based on a simple model of a two-level atom interacting with a
single mode of the cavity field. Qualitatively our results are valid for a real atom
in a physical cavity.}}
\endabstract
\normalsize
\vspace{0.5cm}
\hspace{2cm}
\vspace{0.5cm}


\section{Introduction}
\label{intro}

Much attention was attracted recently in literature to investigation of the dynamical
Casimir effect (DCE), i.e. the process of photon creation in a cavity with variable
length or shape, or a cavity filled with matter, dielectric susceptibility of which
varies due to the action of a strong alternating external electromagnetic field, see
e.g. Refs.\cite{Do,DQP} and citations therein.  Another interesting effect associated
with variation of cavity parameters is modulation of Lamb shift by slow motion of the
cavity walls which was discussed in Refs.\cite{Loz,PRL}. In the present paper we
consider another effect, namely, excitation of an atom, ion or molecule, placed
inside the initially empty non-stationary cavity. The atom, ion or molecule can be either trapped
inside the cavity by external fields, or pass through the cavity in a cold low
intensity beam.

We will show that there exist two mechanisms of excitation of the atom placed in the
initially empty cavity. One of them arises due to the interaction of the atom with
the non-stationary electromagnetic field created by modulation of cavity parameters.
We distinguish three regimes of this mechanism determined by relations between the
characteristic time of variation of cavity parameters $\tau$ and the natural
transition time of the atom $\tau_0$. If $\tau\ll\tau_0,$ the atom does not have time
to follow variation of the field state, and hence the process of its excitation is a
two-stage process. First, Casimir photons are created by the DCE, as if the atom was
absent, and then these photons are absorbed by the atom.
Thus in this regime the atom can be considered as a detector for the final
quantum state of the electromagnetic field inside the cavity.
The second regime, which can
be realized if $\tau \sim \tau_0$, is of more interest. In this case the atom during
the period of time $\tau$ interacts with varying electromagnetic field which cannot
be decomposed into photons on principle. As a consequence, the excitation process
cannot be reduced to absorption of the Casimir photons. If $\tau\gg\tau_0$, we have
the adiabatic regime for which the effect is exponentially small.

Another, and may be {\it the most interesting mechanism, is "shaking" of the atom due to
fast modulation} of its ground state {\it Lamb shift} which takes place as a result of fast
variation of cavity parameters. The last mechanism has no connection with the vacuum
dynamical Casimir effect at all. Moreover, it opens a new channel of photon creation
in the cavity. We will show also that the presence of the {\it atom} in the {\it cavity alters
statistics of the created photons} even if one disregards the shaking effect. It
happens because in the case $\tau\sim\tau_0$ the back reaction of the atom to the
field alters the DCE dynamics and leads to formation of the final photon states
different from those which would have been formed in the absence of the atom. In
particular, the back reaction effect removes the restriction for creation of only
even number of photons and also alters the expectation value for the number of
created photons.

Our consideration is based on a simple model describing a two-level atom interacting
with a single mode of quantized electromagnetic field. The Hamiltonian of the model
can be expressed in terms of the Pauli matrices
$\sigma_3=2\sigma_{+}\sigma_{-}-1$, $\sigma_{\pm}$ and the destruction and creation
Bose operators for the field mode $a$ and $\hc{a}$ \eq{Ham}{
H=E_0\frac{1+\sigma_3}{2}+\omega(t)\hc{a}a+
i\frac{\dot\omega(t)}{4\omega(t)}\left(a^2-{\hc{a}}^2\right)+
\lambda(\sigma_{+}+\sigma_{-})(a+\hc{a}).} Here $E_0\sim\tau_o^{-1}$ is the atomic
transition frequency \footnote{We use units $\hbar=1$.}, $\omega(t)$ is the frequency
of the mode, which depends on time through variable parameters of the cavity, and $\lambda$ is
the coupling constant. Let us remind that in the case of nearly resonant cavity mode
$\omega\approx E_0$ one can omit fast oscillating terms in the interaction
Hamiltonian $\lambda(\sigma_{-}a+\sigma_{+}\hc{a})$ (see e.g. \cite{AE}), and the
model reduces to the so-called generalized Jaynes-Cummings model \cite{Dod}.
Moreover, if $\omega={\rm const}$, then the model reduces to the standard exactly
integrable Jaynes-Cummings model \cite{J,JC}, which describes the interaction of the
two-level atom with a stationary mode of the electromagnetic field. Finally, if
$\lambda=0$, then our Hamiltonian reduces to the one simulating the DCE in the
considered mode of the field \cite{Law}.

Obviously one cannot rely on the one-mode approximation for obtaining relevant
quantitative results even for the resonant case (may be except the trivial case of
the adiabatic limit when no photons are created and atomic excitation is
exponentially damped), see e.g. \cite{Dod}. The reason for that is the strong
interaction between modes in the non-stationary cavity, so that properties of the
"dressed" resonant mode can differ very strongly from those of a single harmonic
oscillator. Nevertheless, we use the one-mode approximation in this paper because
even such a simple model affords one to trace the main qualitative features of the
interaction of the atom and the quantized electromagnetic field inside the
non-stationary cavity.

\section{Atomic excitation. Instantaneous approximation for the generalized
Jaynes-Cummings model} \label{at_DC}

If the atom-field interaction is neglected, i.e. the coupling constant $\lambda$ is
equal to $0$, then for $\omega={\rm const}$ the stationary states of the model
(\ref{Ham}) (bare states) $|n,\da\ket$ and $|n,\ua\ket$ are defined by the conditions
$$ \sigma_3|n,\da\ket=-|n,\da\ket,\quad \hc{a}a|n,\da\ket=n|n,\da\ket,\quad
\sigma_3|n,\ua\ket=+|n,\ua\ket,\quad \hc{a}a|n,\ua\ket=n|n,\ua\ket. $$ For
$\omega\ne{\rm const}$ the Heisenberg-picture solutions read \eq{L=0}{
\sigma_{-}(t)=\sigma_{-}^{(0)}e^{-iE_0t},\quad a(t)=a^{(0)}(t)\equiv\alpha(t)
a_{\inn}+\beta(t) \hc{a}_{\inn},} where the functions $\alpha(t)$ and $\beta(t)$
satisfy the equations \eq{ab_Eqs}{
\dot\alpha=-i\omega\alpha-\frac{\dot\omega}{2\omega}\beta^*,\quad
\dot\beta=-i\omega\beta-\frac{\dot\omega}{2\omega}\alpha^*,\quad
\lim_{t\to-\infty}\alpha(t)e^{i\omega_1 t}=1,\quad\beta(-\infty)=0.}Here $\omega_1$
denotes the initial frequency of the mode, $\sigma_{-}^{(0)}$, $\sigma_{+}^{(0)}$
have the meaning of atomic lowering and raising operators and $a_{\inn}$,
$\hc{a}_{\inn}$ are respectively destruction and creation operators for photons
in the in-region. They serve for definition
of in-states of the system. If the mode frequency approaches the value $\omega_2$
when $t\to+\infty$, the coefficients $\alpha(t)$, $\beta(t)$ acquire asymptotic
behaviour of the form \eq{fin_ab}{ \left[
\begin{array}{l}\disp
\alpha \\ \disp \beta
\end{array}\right]=
\left[
\begin{array}{l}\disp
\alpha_{\infty} \\ \disp \beta_{\infty}
\end{array}\right]\,\cdot\,e^{-i\omega_2 t}, \quad t\to+\infty.}
Thus the destruction operator for $t\to+\infty$ depends on time as
$a(t)=a_{\out}e^{-i\omega_2t}$ and defines the stable out-vacuum as well as
many-photons states. If $\beta_{\infty}\ne 0$ then these states differ from the
corresponding in-states. This indicates creation of photons from vacuum in the
cavity, i.e. the DCE.

Let the state $|0,\da\ket$, describing the atom in the ground state in the absence of
photons in the cavity, be the initial state of the system. Then the expectation value
of the number of created photons is given by the formula \eq{NCE}{ \bar N_{{\rm
DCE}}=\bra 0,\da|\hc{a}_{\out}a_{\out}|0,\da\ket= |\beta_{\infty}|^2.} In particular,
if the frequency $\omega$ at $t=0$ suddenly changes from $\omega=\omega_1$ to
$\omega=\omega_2$, then the number of created photons is given by the simple
expression $\bar N_{{\rm DCE}}=(\omega_2-\omega_1)^2/(4\omega_1\omega_2)$. As it
follows from Eqs.(\ref{ab_Eqs}) and formula (\ref{NCE}), the DCE is determined by the
third term in the Hamiltonian (\ref{Ham}) proportional to $\dot\omega$.

If $\lambda\ne 0$, then the atom interacts with the varying field in the cavity and
can be excited. Let us first consider this effect in the framework of the generalized
Jaynes-Cummings model under assumption that the characteristic time $\tau$ of
variation of the mode frequency $\omega(t)$ is parametrically smaller than all other
parameters in the problem with dimension of time  $\tau\ll
E_0^{-1},\;\omega_1^{-1},\;\omega_2^{-1}$ (instantaneous approximation). It appears
that such a problem admits exact solution and there is no need to use the
perturbation theory expansion with respect to the coupling constant $\lambda$.

If $\omega={\rm const}$, we have the standard Jaynes-Cummings model with the
Hamiltonian $H_{\omega}^{(JC)}=\frac12E_0(1+\sigma_3)+\omega\hc{a}a+
\lambda(\sigma_{+}a+\sigma_{-}\hc{a})$. The exact (dressed) stationary states of this
model are \cite{J}, \eq{dr_st1}{
\begin{array}{l}\disp
|n,\da\ket_{\lambda\omega}=S_n^{(-)}|n,\da\ket-R_n^{(-)}|n-1,\ua\ket,\quad
|n-1,\ua\ket_{\lambda\omega}=R_n^{(+)}|n-1,\ua\ket+S_n^{(+)}|n,\da\ket,\el
R_n^{(\pm)}=\frac{\sqrt{\frac{\Delta^2}{4}+\lambda^2n}\pm
\frac{\Delta}{2}}{\sqrt{\frac{\Delta^2}{2}+ 2\lambda^2n\pm
\Delta\sqrt{\frac{\Delta^2}{4}+\lambda^2n}}},\quad
S_n^{(\pm)}=\frac{\lambda\sqrt{n}}{\sqrt{\frac{\Delta^2}{2}+2\lambda^2n\pm
\Delta\sqrt{\frac{\Delta^2}{4}+\lambda^2n}}},\quad\Delta=E_0-\omega.
\end{array}}
The corresponding energy levels read \cite{J,JC} \eq{dr_levs1}{
E_{n,\da}^{(\lambda)}=\omega n+\frac{\Delta}{2}-
\sqrt{\frac{\Delta^2}{4}+\lambda^2n},\quad\quad E_{n-1,\ua}^{(\lambda)}=\omega
n+\frac{\Delta}{2}+ \sqrt{\frac{\Delta^2}{4}+\lambda^2n}.} Note that the ground state
$|0,\da\ket_{\lambda\omega}=|0,\da\ket$ is not dressed (this is an artifact of the
Jaynes-Cummings model) and does not experience Lamb shift.

The Hamiltonian of the generalized Jaynes-Cummings model can be represented in the
form \eq{H_ns}{ H(t)=H_{\omega_1}^{(JC)}+(\omega(t)-\omega_1)\hc{a}a+
\frac{i\dot\omega}{4\omega}\left(a^2-{\hc{a}}^2\right).} In the instantaneous
approximation the first of additional (in comparison with the standard model)
time-dependent terms in (\ref{H_ns}) is of $\theta$-function type while the second
one is of $\delta$-function type. Hence, according to the general rules of the
instantaneous perturbation theory (see e.g. \cite{LL}) the amplitude of transition
with excitation of the atom and creation of $n$ photons
$|0,\da\ket\to|n,\ua\ket_{\lambda\omega_2}$ is equal to  \eq{ampl_DC}{
A_{n\ua}={}_{\lambda\omega_2}\bra n,\ua|e^{-iW}|0,\da\ket= S_{n+1}^{(+)\,*}\bra
n+1|e^{-iW}|0\ket,} where $W$ is the operator of "percussive" or sudden interaction
$$ W=\frac{i\Theta}2\left(a^2-{\hc{a}}^2\right),\quad \Theta=\frac12\intRR
\frac{\dot\omega}{\omega}\,dt= \frac12\ln\left(\frac{\omega_2}{\omega_1}\right), $$
which arises due to the DCE term in the Hamiltonian (\ref{H_ns}). Note that the
amplitude (\ref{ampl_DC}) is not equal to zero only if $n=2j+1$. Using then an easily
derived formula $$ e^{-iW}|0\ket=\left(\frac{2\sqrt{\rho}}{1+\rho}\right)^{1/2}
\sum\limits_{j=0}^{\infty} (-1)^j\left(\frac{\rho-1}{\rho+1}\right)^j\,
\sqrt{\frac{(2j-1)!!}{2^j j!}}\;|2j\ket,\quad \rho=\frac{\omega_2}{\omega_1},$$ the
following expression for the total probability of atomic excitation can be obtained
\eq{w_DC1}{
\begin{array}{l}\disp
w_{\ua}=\sum\limits_{j=0}^{\infty}|S_{2j}^{(+)}|^2\, |\bra 2j|e^{-iW}|0\ket|^2= \el
=\frac{2\xi^2\sqrt{\rho}(\rho-1)^2}{(1+\rho)^3}
\sum\limits_{j=0}^{\infty}\frac{(2j+1)!!}{2^j j!} \frac{1}{\frac12+4\xi^2 (j+1)+
\sqrt{\frac14+2\xi^2 (j+1)}}\, \left(\frac{\rho-1}{\rho+1}\right)^{2j}.
\end{array}}
We have introduced here the dimensionless coupling constant $\xi=\lambda /\Delta$. It
is amusing that the obtained probability depends only on two parameters
$\rho=\omega_2/\omega_1$ and $\xi=\lambda/\Delta$, though {\it a priori} there exist four
(if $\tau\to 0$) independent parameters $E_0$ , $\omega_1$, $\omega_2$ and $\lambda$
with dimension of energy and one could suppose that $w_{\ua}$ depended on three their
dimensionless combinations. This result is probably an artifact of the model. Let us
also note that the probability of atomic excitation does not change under the
substitution $\rho\to\rho^{-1}$. It means that the probability depends only on the
ratio of the initial and final frequencies  but does not depend on whether the
frequency has grown or decreased.

In the weak coupling limit $\xi\ll 1$ we have \eq{w_DC2}{ w_{\ua}\approx \xi^2 \bar
N\left\{1-\frac{6\xi^2}{\bar N+1} \left[1+\frac{\bar N}{2(\bar N+1)}\left(1-3(\bar
N+1)^{-5/4}\right) \right]\right\},\quad \bar N=\frac{(\rho-1)^2}{4\rho},} (here the
quantity $\bar N$ has the meaning of average number of photons created in the absence
of the atom), while in the strong coupling limit (or resonance) $\xi\gg 1$ one
obtains \eq{w_DC3}{ w_{\ua}\approx \frac{(\rho-1)^2}{2(1+\rho)(1+\sqrt{\rho})^2}+
O\left(\xi^{-1}\right)=\frac{\bar N}{2(1+\bar N+\sqrt{\bar N+1})}.} Note that in the
strong coupling limit $w_{\ua}\approx 1/2$ for $\rho\to 0$ or $\rho\to\infty$ ($\bar
N\to\infty$ in both cases), i.e. the atom excitation is the most efficient for this
range of parameters within the framework of the model. The behaviour of
$w_{\ua}(\rho,\xi)$ in the intermediate range of parameters is shown in Fig.
\ref{Fig1}.

It is easily seen that the first term on the right-hand side of Eq.(\ref{w_DC2}) is
equal to the product of the average number of photons $\bar N$ created due to the DCE
and the probability $(\lambda/\Delta_2)^2$ of absorption of a single photon by the
atom. It means that excitation of the atom in the weak coupling limit of the
instantaneous approximation occurs due to the trivial process of absorption of
Casimir photons created by the DCE. The expression (\ref{w_DC1}) for the probability
at arbitrary $\xi$ does not have such a simple structure. Nevertheless, the main
conclusion remains true. It immediately follows from Eq.(\ref{ampl_DC}) for the
transition amplitude. Indeed, it is easy to see that the ket-vector
$e^{-iW}|0,\da\ket$ which enters the amplitude (\ref{ampl_DC}) coincides with that
initial state of the system which would have arisen if the atom was placed into the
cavity \tit{after} the cavity had become stationary and the DCE had taken place.
Therefore, in the framework of the instantaneous approximation for the generalized
Jaynes-Cummings model the process of atomic excitation inside the initially empty
non-stationary cavity reduces to absorption of Casimir photons. This conclusion is
confirmed also by the expression for the average number of photons created in the
presence of the atom, which can be easily derived for arbitrary $\lambda$. It reads
\eq{n_DC}{ \bar n=\sum\limits_{n=0}^{\infty}
n\left(|A_{n\da}|^2+|A_{n\ua}|^2\right)=\bar N_{{\rm DCE}} -w_{\ua},} where the
amplitude $A_{n\da}$ for creation of $n$ photons without atomic excitation is
determined by the expression similar to Eq.(\ref{ampl_DC}). More complicated form of
the probability (\ref{w_DC1}) in comparison with the first term of the expansion
(\ref{w_DC2}) is explained by creation of out-photons strongly coupled to the atom
rather than free photons for not small values of the coupling constant.

Of course, it is not surprising  from the physical point of view that the process of
atomic excitation reduces to the two-stage process in the instantaneous approximation
since in the case $\tau\ll E_0^{-1}$ the atom feels changes of the field state only
after the cavity has become stationary and the DCE has taken place. However, we will
demonstrate in the next sections that essentially new effects arise if the limits of
the used approximation or the considered model are exceeded.

\section{Non-Casimir atomic excitation}
\label{at_exc}

We will discuss now two extensions of the simple model considered in the preceding
section. In both of them excitation of the atom cannot be reduced to absorption of
photons created by the DCE.

First, we will study atomic excitation in the framework of the generalized
Jaynes-Cummings model for the case when the characteristic time of variation of the
mode frequency is of the order of the atomic transition time $\tau \sim E_0^{-1}$.
Following the standard methods of the time-dependent perturbation theory \cite{LL},
for time evolution of the initial state in the first order with respect to $\lambda$
we get \eq{dyn_t}{ |0,\da\ket_{\lambda,t}=|0,\da\ket-i\lambda {\cal B}(t)\,|1,\ua\ket
+O\left(\lambda^2\right),\quad {\cal B}(t)=\intT \beta(t') e^{iE_0t'}\,dt'.} After
performing integration by parts we obtain for $t\to+\infty$ \eq{B}{ {\cal
B}(t)=\frac{-i}{\Delta_2}\left\{ \beta_{\infty}(\tau)e^{i\Delta_2 t}- \intRR
dt'\,e^{i\Delta_2 t'}\,\frac{d}{dt'} \left[\beta(t')e^{i\omega_2 t'}\right]\right\}.}
We see from Eq.(\ref{B}) that the the long-time asymptotic expression for the
transition amplitude $-i\lambda {\cal B}(t)$ consists of two terms one of which is
constant while the other oscillates with the frequency $\Delta_2= E_0 - \omega_2.$
According to the general rules of the time-dependent perturbation theory the
oscillating term has no relation to the transition probability and hence should be
omitted \cite{LL}. Then the probability of atomic excitation can be represented in
the form \eq{w_tau}{ w_{\ua}=\frac{\lambda^2}{\Delta_2^2} |\beta_{\infty}(\tau)|^2
\cdot F(\tau), \quad F(\tau)=\left\vert\intRR dt'\,e^{i\Delta_2 t'}\,\frac{d}{dt'}
\left[\frac{\beta(t')}{\beta_{\infty}}e^{i\omega_2 t'}\right] \right\vert^2.} Here
the dimensionless function $F(\tau)$ determines efficiency of atomic excitation in
the cavity as compared with the channel of excitation by $\bar N$ Casimir photons
($\bar N =|\beta_{\infty}(\tau)|^2$ ) created by the DCE. As it should be, we have
$F(0)=1$, and Eq.(\ref{w_tau}) at $\tau=0$ is in agreement with Eq.(\ref{w_DC2}) for
the probability of atomic excitation derived in the weak coupling limit of the
instantaneous approximation.

>From the physical point of view the values $F\approx 1$ correspond to the regime of
atomic excitation due to absorption of Casimir photons, while the values $F< 1$ or
$F> 1$ indicate that excitation occurred due to the interaction of the atom with
essentially non-stationary field during the transient process when the out-states of
Casimir photons had not been formed yet. It follows from the numerical calculations
that, at least for the investigated values of parameters, the function $F(\tau)$
increases monotonously with $\tau$, so that we have the estimation $F(\tau)>1$ for
$\tau>0$. Moreover, for some values of parameters the excitation efficiency $F$
becomes very large ($\sim10$, or even $\sim10^2$) if $\tau\sim E_0^{-1}$. Therefore,
in spite of fast (exponentially fast in the adiabatic limit $\tau\to\infty$) decrease
of average number of the created Casimir photons with $\tau$, the probability of
atomic excitation for the case $\tau\sim E_0^{-1}$ can become greater than the one
for the instantaneous case $\tau=0$ (see Fig. \ref{Fig2}).

To understand this effect, let us consider the behaviour of the "instantaneous number
of photons" $|\beta(t)|^2$ with time, which is shown in Fig. \ref{Fig3}. It is seen
from Fig.\ref{Fig3} that for $\tau>0$ $|\beta|^2$ achieves its maximum value in the
transient region. Moreover, in this region $|\beta|^2\gg|\beta_{\infty}|^2$ and therefore
contribution of this region results in the effect of amplification of the excitation
probability. It is clear that for essential amplification of the probability the
atomic transition time $\tau_0 \sim E_0^{-1}$ must be of the order of $\tau$. This
explains the resonance-like shape of the curve $w_{\ua}(\tau)$ in Fig. \ref{Fig2}
(we suppose at this point that $\omega_1,\;\omega_2\sim E_0$).

Let us now return to the instantaneous approximation but for the model with the
Hamiltonian (\ref{Ham}) which differs from the Jaynes-Cummings Hamiltonian by the
terms $\lambda(\sigma_{-}a+\sigma_{+}\hc{a})$. The model with these terms in the
Hamiltonian ceases being exactly soluble and we will use perturbation theory for its
analysis. If $\omega={\rm const}$, the corrections to the state vectors corresponding
to the stationary states of the system are of the first order \eq{dr_sts}{
\begin{array}{l}\disp
|n,\da\ket_{\lambda\omega}=|n,\da\ket+\frac{\lambda\sqrt{n}}{\omega-E_0}
|n-1,\ua\ket-\frac{\lambda\sqrt{n+1}}{\omega+E_0}|n+1,\ua\ket
+O\left(\lambda^2\right),\el
|n,\ua\ket_{\lambda\omega}=|n,\ua\ket+\frac{\lambda\sqrt{n}}{\omega+E_0}
|n-1,\da\ket-\frac{\lambda\sqrt{n+1}}{\omega-E_0}|n+1,\da\ket
+O\left(\lambda^2\right),
\end{array}}
while shifts of the bare energy levels are of the second order with respect to the
coupling constant $\lambda$ \eq{dr_levs}{
E_{n\da}^{(\lambda)}=\left(\omega+\frac{2\lambda^2E_0}{\omega^2-E_0^2}
\right)n-\frac{\lambda^2}{\omega+E_0}+O\left(\lambda^3\right),\quad
E_{n\ua}^{(\lambda)}=\left(\omega-\frac{2\lambda^2E_0}{\omega^2-E_0^2}
\right)n+E_0-\frac{\lambda^2}{\omega-E_0}+O\left(\lambda^3\right).} The second order
terms in Eqs.(\ref{dr_levs}) which are not proportional to "$n$" can be interpreted
as Lamb shifts of atomic levels. Note that in contrast to the Jaynes-Cummings model
the ground state of the system $|0,\da\ket_{\lambda\omega}$ is also dressed and is
characterized by the Lamb shift $E_L=-\lambda^2/(\omega+E_0)$.

The excitation probability can be calculated according to the same algorithm which we
have used in the preceding section but with the state vectors (\ref{dr_sts}) instead
of (\ref{dr_st1}).  Let us split the amplitude $A_{n\ua}$ for excitation of the atom
and creation of $n$ photons into two parts, \eq{ampl_split}{
\begin{array}{l}\disp
A_{n\ua}={}_{\lambda\omega_2}\bra n,\ua|
e^{-iW}|0,\da\ket_{\lambda\omega_1}=A_{n\ua}^{(L)}+A_{n\ua}^{(C)}, \quad
A_{n\ua}^{(L)}={}_{\lambda\omega_2}\bra n,\ua|0,\da\ket_{\lambda\omega_1}, \el
A_{n\ua}^{(C)}={}_{\lambda\omega_2}\bra n,\ua|
(e^{-iW}-1)|0,\da\ket_{\lambda\omega_1}.
\end{array}}
Such splitting of the amplitude does not lead to extraction of the
non-resonant effects associated with the new terms in the Hamiltonian. These effects
contribute of course to both parts of the excitation amplitude (\ref{ampl_split}).
However, the non-resonant effects in the second term can be considered as small
corrections. At the same time the first term, which corresponds to the parametric shaking
of the atom, originates entirely from those nonresonant effects in the second
term. Note that such shaking excitation of the atom {\it due to the fast modulation
of the Lamb shift} (see below) is akin to Migdal shaking of the atom due to
$\beta$-decay \cite{LL}.
This term was absent in the
amplitude (\ref{ampl_DC}) because the state $|0,\da\ket$ was the exact ground state
of the system in the Jaynes-Cummings model for arbitrary values of the mode frequency
and thus was orthogonal to all excited states. Moreover, the second term in the sum
(\ref{ampl_split}) is defined by the instantaneous change of the ground state
$(e^{-iW}-1)|0,\da\ket_{\lambda\omega_1}$ of the system due to the DCE and would
disappear if we have excluded the DCE term proportional to $\dot\omega$ from the
Hamiltonian (\ref{Ham}), i.e. formally have put the operator $W$ to zero. At the same
time, the term $A_{n\ua}^{(L)}$ would have remained unchanged after such a procedure.
This shows that this term corresponds to an essentially new effect which has no
connection with the DCE at all.

Using Eqs.(\ref{dr_sts}), it is easy to find the contribution of the shaking effect
to the probability of excitation of the atom. In the first not vanishing order of the
perturbation theory it is given by the equation \eq{w_nr}{
w_{\ua}^{(L)}=\sum\limits_{n=0}^{\infty}|A_{n\ua}^{(L)}|^2= |{}_{\lambda\omega_2}\bra
1,\ua|0,\da\ket_{\lambda\omega_1}|^2=
\lambda^2\left(\frac1{\omega_2+E_0}-\frac1{\omega_1+E_0}\right)^2= \left(\frac{\delta
E_L}{\lambda}\right)^2,} where $\delta E_L$ denotes variation of the Lamb shift of
the ground state of the atom. It follows from Eq.(\ref{w_nr}) that atomic excitation
due to the shaking effect is accompanied by creation of a photon (or, as it can be
simply shown, greater but always odd number of photons in the higher orders of the
perturbation theory) and arises  due to modulation of the ground state Lamb shift in
the non-stationary cavity. Note that proportionality of the excitation probability to
the squared variation of the ground state Lamb shift (provided that its relative
change is small) is not a specific feature of the considered simple model. It arises
also in completely realistic 3D problem about a real atom placed into a cavity with
variable parameters. The reason is that both the variation of the Lamb shift and the
shaking amplitude are proportional to the first power of the small variation of the
cavity parameter, while the excitation probability is given by the square of the
amplitude. This argument affords one to reproduce Eq.(\ref{w_nr}) for our model
correct at least to a factor. Indeed, in the first order of perturbation theory the
probability of excitation $w_{\ua}^{(L)}$ is proportional to the square of the
coupling constant $\lambda$. Taking into account that $w_{\ua}^{(L)} \sim \delta
E_L^2$, and the variation of the Lamb shift itself $\delta E_L \sim \lambda^2$, we
immediately get $w_{\ua}^{(L)} \sim (\delta E_L/\lambda)^2$ in full agreement with
Eq.(\ref{w_nr}). From this point of view the absence of the shaking mechanism of the
atom excitation in the Jaynes - Cummings model is explained by zero Lamb shift of the
atomic ground state in it.

\section{Back reaction of the atom to the DCE}
\label{back_reac}

It is clear from the preceding sections that the atom placed inside the cavity alters
statistics of created photons. This statement is true for all considered mechanisms
of photon creation and arises from the fact that presence of the atom in the cavity
opens a new channel of the process with excitation of the atom and creation of odd
number of photons, while in the absence of the atom in the cavity only even number of
photons can be created, see e.g. \cite{Do}. This effect is a manifestation of back
reaction of the atom on dynamics of the DCE. In this section we will investigate
influence of the atom on the average number of created photons.

The number of created photons in the instantaneous approximation for the generalized
Jaynes-Cummings model is given by Eq.(\ref{n_DC}). Since the atom in this model is
excited by a single Casimir photon this result is trivial and Eq.(\ref{n_DC}) can be
easily received without special calculations $$ \bar n = N_{\rm DCE}(1-w_{\ua})
+(N_{\rm DCE}-1)w_{\ua}= N_{\rm DCE}-w_{\ua}.$$ As it was already mentioned, the atom
excitation by the shaking effect is also accompanied by creation of photons. The
average number of photons created by this process, however, is of the order of
$(\delta E_L/\lambda)^2$ and is always small in the weak coupling limit. The
influence of finiteness of the characteristic time of the mode frequency variation
could be more efficient since it includes resonant effects. Therefore in the rest of
the section we will consider this aspect of the problem in the framework of the
generalized Jaynes - Cummings model.

Let us consider the operator \eq{Int_Mot}{ N=\hc{a}a+\frac12(1+\sigma_3),} which is
known to be an integral of motion for the stationary Jaynes - Cummings model
\cite{JC}, i.e. in the case $\omega={\rm const}$. If the cavity is not stationary,
time dependence of the operator $N$ is ruled by the equation \eq{dot_N}{ \dot
N=i[H,N]=-\frac{\dot\omega}{2\omega}(a^2+{\hc{a}}^2)\ne 0.} Using
Eqs.(\ref{Int_Mot}), (\ref{dot_N}) and the standard perturbation theory technique, it
is easy to show that the expectation value $\bar N(t)=\bra 0,\da|N|0,\da\ket$
up to the first order with respect to $\lambda$ equals to $\bar
N(t)=|\beta(t)|^2+o(\lambda)$ and, in particular, \eq{1_order}{ \bar N_{\infty}=\bar
N(+\infty)=|\beta_{\infty}|^2+o(\lambda)= \bar N_{{\rm DCE}}+o(\lambda).} Since on
the other hand $\bar N_{\infty}=\bar n+w_{\ua}$ where $\bar n=\bra \hc{a}a\ket$ is
the average number of created photons, we see that with accuracy up to the first
order with respect to $\lambda$ the number of photons created in the presence of the
atom is determined by Eq.(\ref{n_DC}). It could seem therefore that the number of
created photons changes only due to absorption of them by the atom as it happens in
the instantaneous approximation. However as it was shown in section \ref{at_exc}, the
process of atomic excitation is not reduced to absorption of created photons if
$\tau\ne 0$. So that, $w_{\ua}$ may not coincide with the probability of absorption
of a photon. Besides, in the second order with respect to $\lambda$ there appears a
correction to Eq.(\ref{1_order}) which is of the form \eq{dN}{ \delta\bar N(t)=\bar
N(t)-|\beta(t)|^2= 2\lambda^2\left\{|\alpha(t)|^2|{\cal B}(t)|^2-2\,{\rm Re}\,
\left(\alpha(t)\beta^*(t) \intT dt'\,{\cal B}(t')\alpha^*(t')e^{-iE_0t'}\right)\right\}.}
An important point is that the corresponding correction $\delta\bar
N_{\infty}=\delta\bar N(+\infty)$ to the r.h.s. of Eq.(\ref{n_DC}) is of the same
order as the term $w_{\ua}$ and hence both corrections to the number of created
photons are comparable. It means that the first non-vanishing correction to the
number of created photons is different from that one arising due to the Casimir
mechanism.

To consider only that part of the effect which is associated with the contribution
$\delta\bar N$, let us introduce the notation \eq{eta}{ \bar N_{\infty}=\bar N_{{\rm
DCE}}\left(1+ \frac{\lambda^2}{E_0^2}\eta(\tau)\right).} Then the dimensionless
parameter $\eta(\tau)$ characterizes the intensity of the back reaction of the atom
to the DCE. Dependence of the parameter $\eta$ on $\tau$ was studied numerically and
is represented in Fig. \ref{Fig4} for some typical set of the model parameters. As it
is seen from the figure, $\eta(\tau)\propto\tau^2$ for $\tau\ll E_0^{-1}$ and
$\eta(\tau)\sim\tau$ for $\tau\gg E_0^{-1}$. The initial quadratic behaviour of the
dependence $\eta(\tau)$ can be explained by coherence of the interaction of the atom
with two Casimir photons emitted simultaniously (after delay $\sim\tau\ll E_0^{-1}$).
The linear dependence in the adiabatic regime arises due to loss of the coherence
between these subsequent photon emissions (since the delay $\sim\tau$ becomes
greater than $E_0^{-1}$).
                         
To conclude, let us note that the considered effects of the interaction of atom with
quantized field inside a non-stationary cavity could be realized experimentally for
example by passing an atomic beam through a microwave cavity. An impurity in a
transparent cavity can serve another realization. Optical properties of the cavity
can be rapidly varied by ultra-short laser pulses \cite{Loz1}. The time of variation of optical
properties of the resonator in such experiments can be easily made of the order of
the inverse of atomic frequency.

\vspace{0.5 cm}

The authors wish to thank V.D. Mur for helpful discussions. This work was supported
by the Russian Foundation for Basic Research (grant 00-02-16354).

\vspace{1cm}

\parbox{15cm}{
\parindent=3cm
\hrule

\parindent=0.5cm}

\newpage

\centerline{{\bf \uppercase{Captures for the Figures.}}}

\begin{figure}[hhhhh]
\caption{Atom excitation probability $w(\rho,\xi)$ in the instantaneously
non-stationary cavity in the Janes - Cummings model  for $\rho>1$;
$\rho=\omega_2/\omega_1$, $\xi=\lambda/\Delta$; $\omega_{1,2}$ are initial
and final frequencies of the mode, $\lambda$ is coupling constant and
$\Delta$ is detuning parameter.}
\label{Fig1}
\caption{Influence of finiteness of the duration of transient region on the atom
excitation probability. Calculations has been performed for the model dependence
$\omega(t)=(\omega_1+\omega_2 e^{t/\tau})/(1+e^{t/\tau})$,
parameters used: $E_0=0.8,\;\omega_1=0.5,\;\omega_2=5.0$. $E_0$ is the
atomic transition frequency.}
\label{Fig2}
\caption{Typical time evolution of the quantity $|\beta(t)|^2$,
$\beta$ is Bogolubov coefficient; parameters used:
$\tau=1.0,\;\omega_1=0.5,\;\omega_2=5.0$.}
\label{Fig3}
\caption{Relative correction to the average number of created
photons due to interaction with the atom; parameters used:
$E_0=0.8,\;\omega_1=0.5,\;\omega_2=5.0$.}
\label{Fig4}
\end{figure}
                                                        
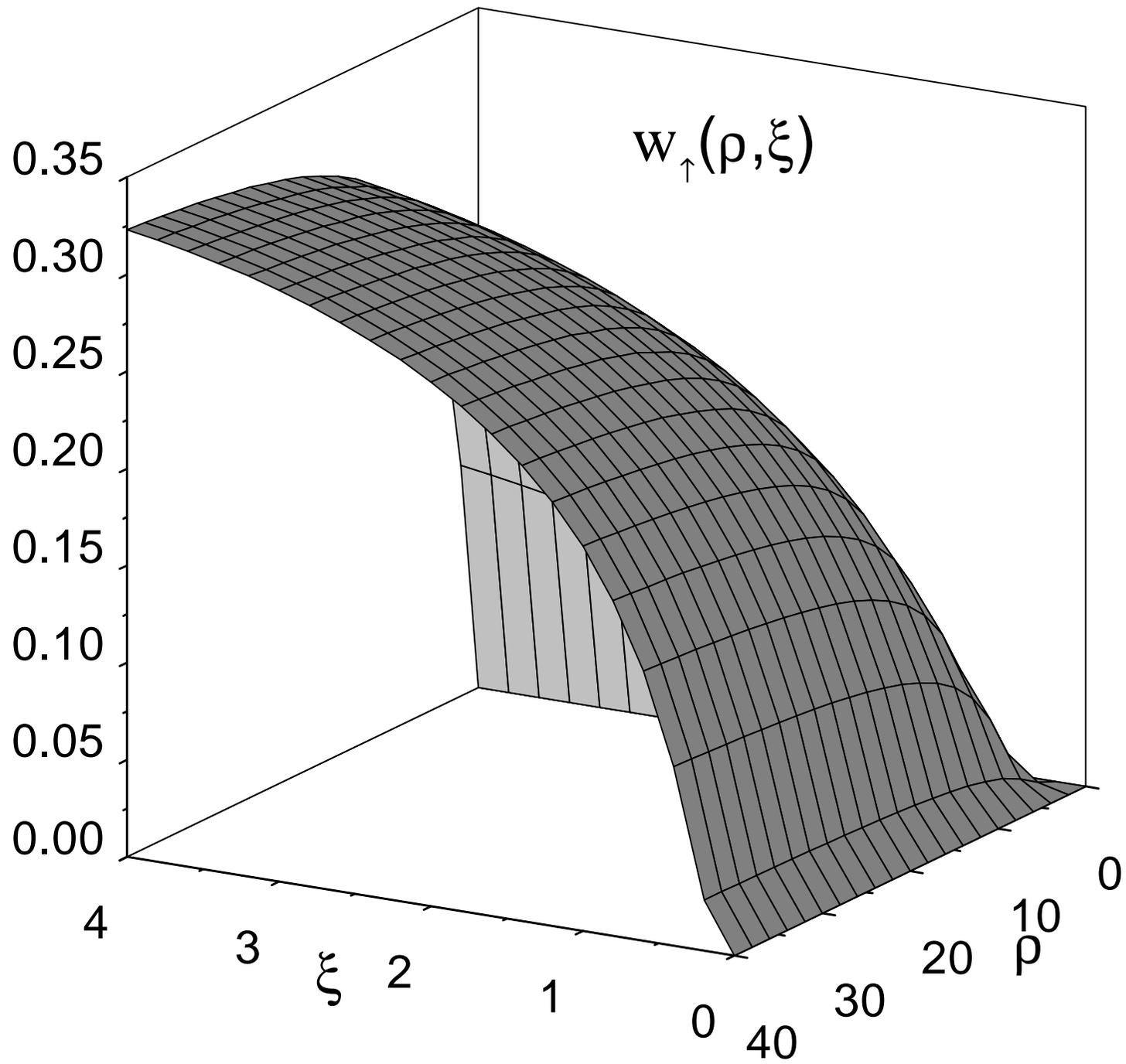
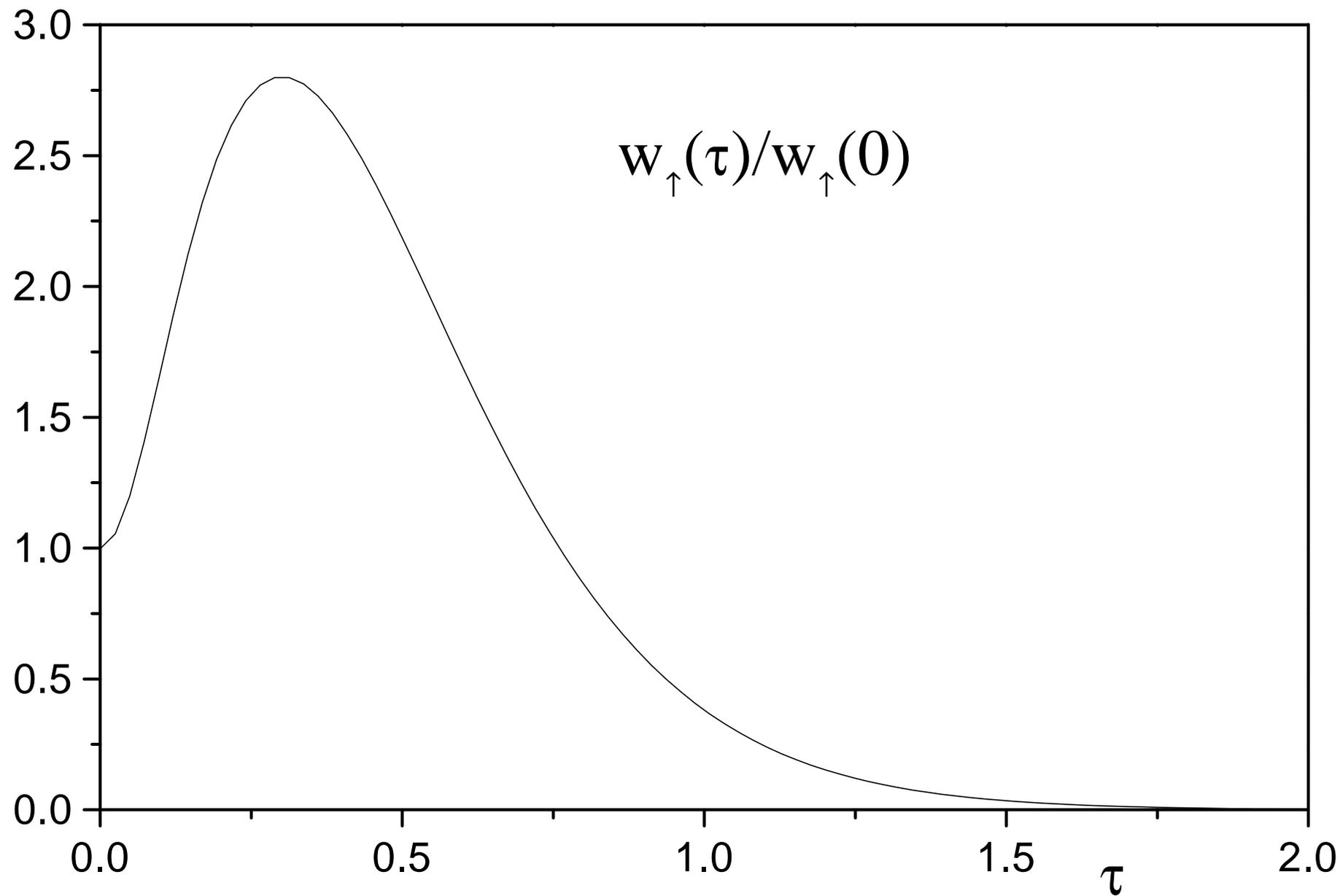
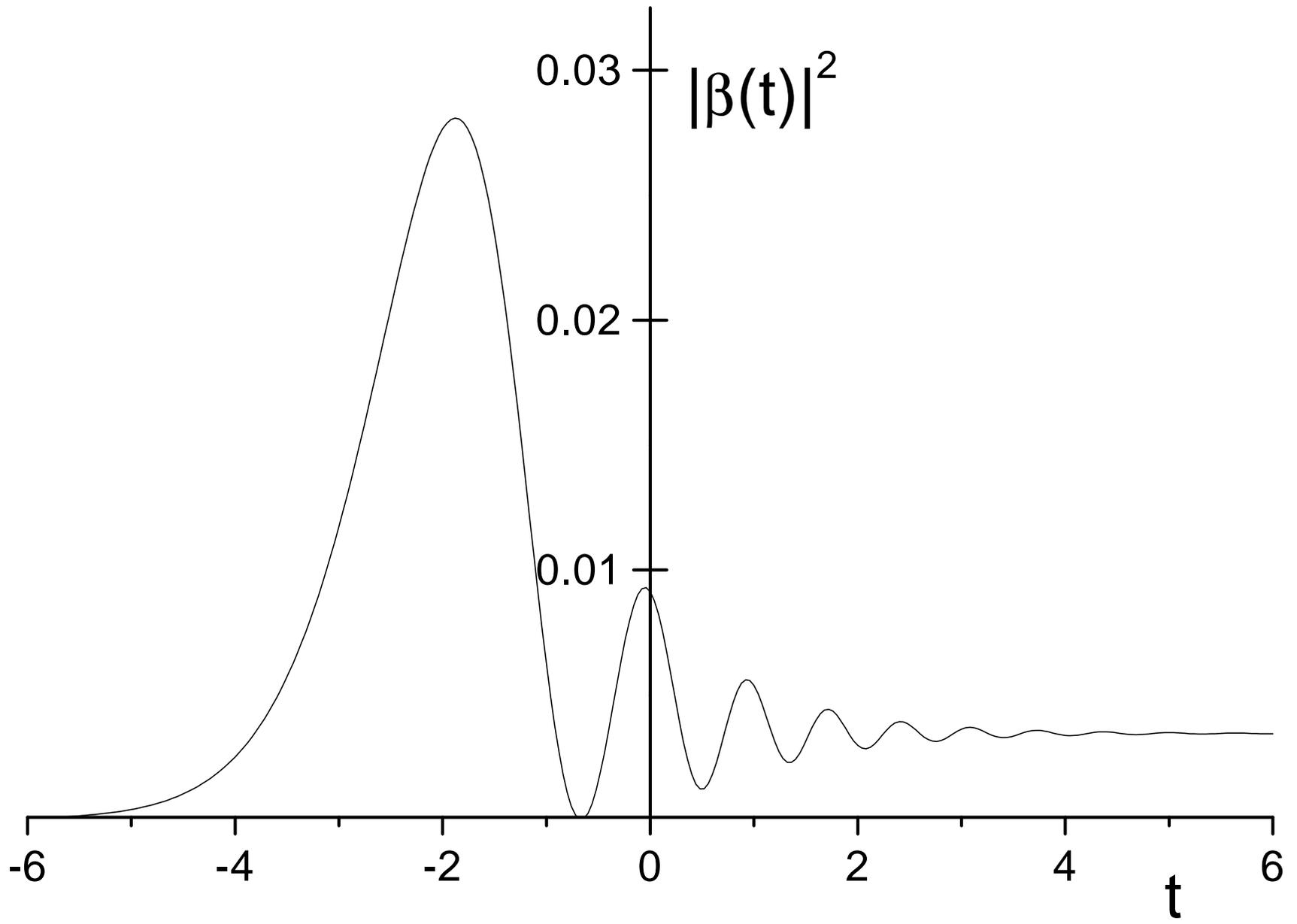
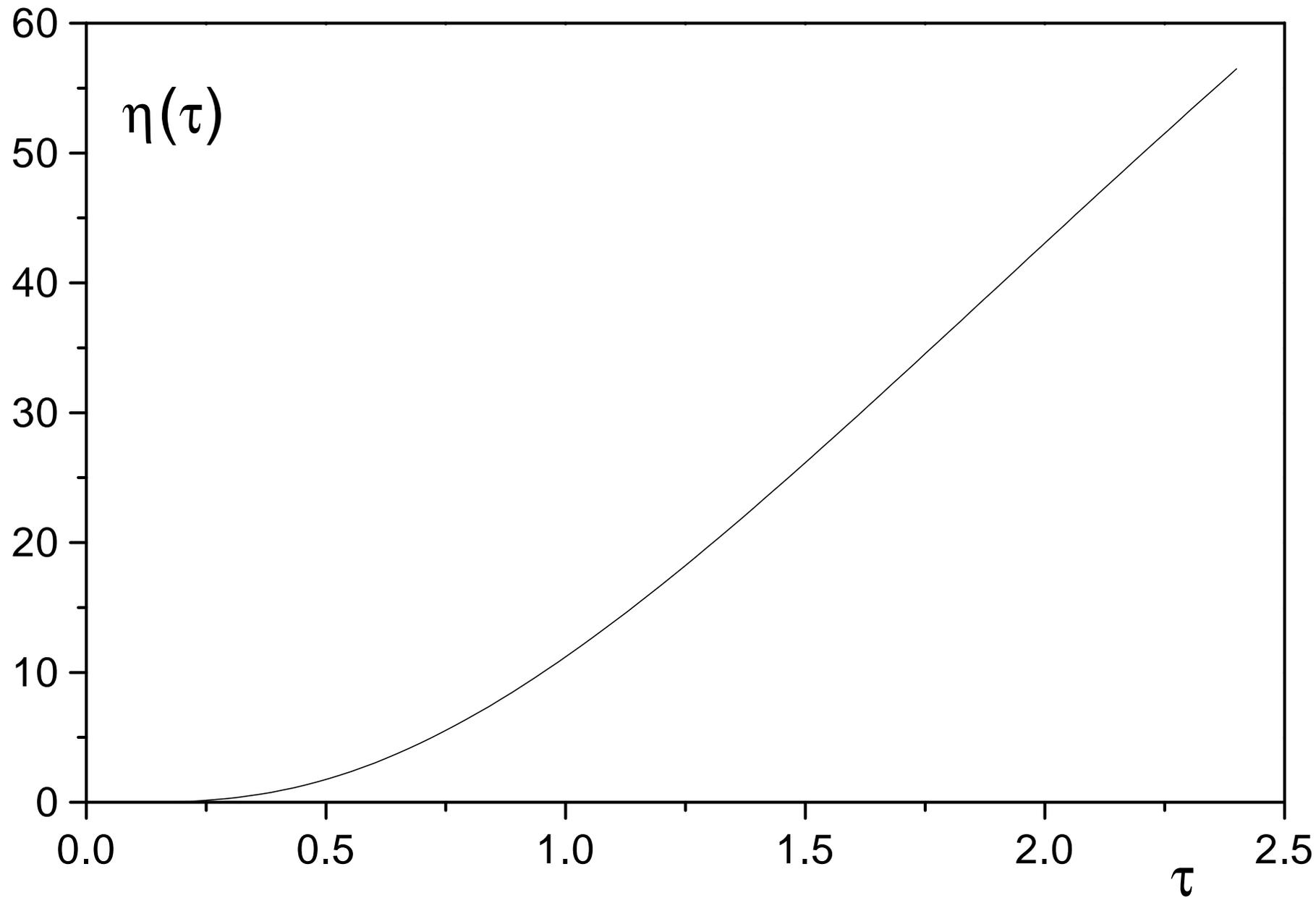
\end{document}